\documentclass{article}
\pdfoutput=1

\usepackage{graphicx}
\usepackage{subcaption}

\usepackage[english]{babel}

\usepackage[letterpaper,top=2cm,bottom=2cm,left=3cm,right=3cm,marginparwidth=1.75cm]{geometry}

\usepackage{amsmath}
\usepackage{graphicx}
\usepackage[colorlinks=true, allcolors=blue]{hyperref}

\title{\textbf{Academic Miscommunication in Physics}}
\author{\textbf{Scott C. Scharlach$^1$}}

\begin{document}

\maketitle

\begin{center}
{\textit{$^1$Pomona College \\
333 N College Way \\
Claremont, CA 91711, USA}}
\end{center}

\begin{abstract}

Academic Miscommunication (AM) is the phenomenon of a professor’s expectations, beliefs, or goals in the classroom differing from those of the students. In this study, a survey was given to undergraduates in two introductory physics classes (one class intended for majors, another for non-majors) who had recently received their grade for the first exam of the semester. The survey measured whether the exam had matched the students’ expectations of the exam, both in form and in content. From these responses, students were assigned an “AM score,” a numerical value in which higher numbers indicate a greater deviation between the actual exam and the students’ expectations. Students’ exam scores were compared with their AM scores; both classes displayed an inverse linear correlation between exam scores and AM scores. The class for majors displayed $R^2 = 0.291$ (n=14), corresponding to a p-value of 0.031. The class for non-majors displayed an $R^2 = 0.457$ (n=30), corresponding to a p-value of 0.000035. The survey also inquired about several other phenomena, such as Impostor Syndrome and the growth-versus-fixed mindset paradigm; 20 graphs were plotted in total. The Bonferroni correction therefore requires a p-value of 0.0025 for statistical significance. We reject the null hypothesis concerning exam scores and feelings of AM for the physics class for non-majors. In contrast, we fail to reject the null hypothesis concerning exam scores and feelings of AM for the physics class for majors. Our findings indicate that non-physics-major students who performed poorly on the first exam of the semester had misunderstood the expectations about the style and content of the exam.

\end{abstract}

\section{Introduction}

\subsection{Physics Education Research (PER)}
Physics Education Research (PER) is the subfield of physics which studies how students learn physics. In other words, PER is the psychology of physics students. The subfield also studies how professors can improve their teaching skills and the variety of teaching methods which professors can adopt to achieve this. The core goal of PER is to structure the physics classroom so that it produces students who deeply understand the material and who feel welcomed in the physics community.

Lillian McDermott of the University of Washington originally formalized PER as a recognized subfield of physics in the 1970s. Since then, the field has expanded to many universities, particularly at the University of Colorado at Boulder, Tufts University, and University of Maryland College Park. The online database PER Central lists no less than ten countries with universities that conduct PER, with ninety universities in the United States alone. The number of researchers in the PER community is steadily increasing.

PER serves an essential role in physics because it reveals crucial but counterintuitive truths about undergraduate physics courses. Deslauriers 2019 analyzed a lecture-based course and an active-learning-based course which taught identical material. The study asked students to self-report the level of learning they acquired from the class. The students in the lecture-based class reported feelings of learning which were higher than those reported in the active-learning-based class. However, students in the active-learning-based class consistently outperformed students in the lecture-based class on exams. This result was highly unexpected: students who felt as if they had learned more than their peers in fact learned less than their peers.

The study represents one of many bizarre but important phenomena which PER has discovered. Studies in PER reveal that the art of teaching physics to students is far more complex and nuanced than one might initially suppose. A professor cannot determine whether a student is learning simply by asking him or her. PER helps identify which aspects of a physics course are sub-optimal and how professors might improve the course. PER has the potential to improve the physics classroom by fostering better understanding between the professor and students.

\subsection{Academic Miscommunication (AM)}

This study coins the term “Academic Miscommunication” (AM) to denote the phenomenon in which a student interprets the professor’s words, assignments, or other aspects of the curriculum in a manner that deviates from what the professor had intended. AM is a potentially consequential phenomenon in academia, as an intelligent and hard-working student may perform poorly on exam if the professor and student miscommunicated about what the exam would cover. AM is especially dangerous because it can occur without either the professor or student detecting it until after the student performs poorly on an exam.

Despite the importance of AM, there are very few published PER articles which address the topic directly. Many PER studies focus on students’ misunderstanding of a particular topic in physics, such as experimental uncertainty (Pollard et al. 2018) or negative numbers (Fuadiah et al. 2019). However, these studies focus on a student’s misconception of a topic in physics rather than miscommunication with the professor about expectations in the physics classroom. These papers focus on students misinterpreting the physical concepts themselves, not the intentions or instructions of the teacher.

Crago et al. 1997 investigates miscommunication in the classroom, but the phenomenon is explored at the linguistic level. The study found that students who spoke a different dialect from the teacher, who spoke the teacher’s language as a second language, or who were raised in a different culture from the teacher were more likely to misunderstand the teacher’s words. The study does not explore whether AM can occur even when there is no linguistic barrier between teacher and student.

Titsworth et al. 2015 conducted two meta-analyses comparing the clarity of a teacher's instructions and students' academic performances. The study found that clear instructions lead to higher academic performance, but that teacher clarity only accounts for approximately $\%$13 of the variance in student grades. In other words, clear instructions leads to only a small to moderate increase in student learning. The study does not, however, investigate the phenomenon of students' sometimes performing poorly on an exam even when the student reported that the instructions were clear.

The book \textit{How Learning Works} by Ambrose et al. 2010 states several real-world anecdotes from professors recounting situations involving miscommunication with students. The book explores methods of counteracting miscommunication, but it does not conduct original research regarding the prevalence or causes of AM.

Olson 1990 explores an anecdote from the author's own life regarding miscommunication in the classroom. In particular, the schoolteacher, her daughter (the student), and the author (the parent of the student) each possessed a distinctly different philosophy regarding the purpose of grades and how students ought to be evaluated. Although this paper explores AM directly, it does not examine AM through a systematic survey or qualitative interviews.

The purpose of this study is to investigate the prevalence of AM in the physics classroom and discover whether other social phenomena, such as Impostor Syndrome, novice-versus-expert thinking, and methods of studying correlate with levels of AM. Ultimately, the study aims to provide professors with a tool for predicting which students may be vulnerable to miscommunication so that the professors may clarify their expectations before the student begins to struggle in the classroom.

\section{Theory and Background}

\subsection{Novice-versus-Expert Thinking}

Redish et al. 1998 demonstrated that experts in physics think about the field differently from novices. The study gave a survey to experts and novices asking about their attitudes, expectations, or perceptions about physics. Redish et al. found that that experts and students deviated significantly in their beliefs about physics. Furthermore, Redish et al. found that novice’s beliefs deviated from experts more strongly at the end of the semester compared to the beginning. In other words, the physics class correlated with a decrease in expert-like thinking among students.

Adams et al. 2004 conducted similar research at the University of Colorado at Boulder with the Colorado Learning Attitudes about Science (CLASS) survey. The study found that physics experts have a philosophical worldview about physics that differs greatly from novices.

Perkins et al. 2005 found that physics courses which focused on expert-like thinking led to greater conceptual student learning than traditional courses. Students in the courses oriented around expert-like thinking were also more likely to take physics courses in the future than their peers.

\subsection{Growth-versus-Fixed Mindset}

The social psychologist Dr. Carol S. Dweck hypothesized that, generally speaking, humans perceive success in one of two formats: the “growth mindset” and the “fixed mindset.” The growth mindset is the belief that humans can develop skills through effort, practice, and perseverance. The fixed mindset is the belief that skills are an unchanging aspect of a human’s identity and that effort has an inconsequential impact on one’s abilities. Dr. Dweck specifies that no human has an entirely growth-oriented mindset or fixed-oriented mindset, but that all humans embody a mixture of both mindsets in different contexts. Dr. Dweck popularized this paradigm in her 2006 book Mindset: The New Psychology of Success.

Mangels et al. 2006 examined students’ responses to hearing feedback on a recent exam. Participants with a fixed mindset tended to pay attention to whether their answers were correct, but they paid little to no attention to the constructive feedback intended to help them learn for the future. In contrast, participants with a growth mindset paid close attention to the constructive feedback, which aided them in learning the previously misunderstood material.

Blackwell et al. 2007 studied middle school students with initially equal academic performance but with differing mindsets. As the curriculum increased in difficulty, students with a growth mindset steadily increased their academic performance, while students with a fixed mindset began to perform more poorly. This research suggests that a students’ mindset is an important dimension which teachers ought to be aware of in the classroom.

\subsection{Impostor Syndrome}

Impostor Syndrome is the feeling of dis-belonging or inadequacy among members of high-achieving communities, even if the individual experiencing Impostor Syndrome is exceptionally talented. Those who feel Impostor Syndrome often feel that they do not deserve to be a member of their community, that their peers will discover their perceived inadequacies, or that they are an impostor masquerading as a talented individual — even if the individual is no less accomplished than their peers. Impostor Syndrome is often accompanied by anxiety, shame, and other negative emotions.

Woolston $\&$ Chris 2016 and Chrousos et al. 2020 reported that Impostor Syndrome affects graduates of Ivy League schools, professional scientists who receive research funding from NASA, theoretical physicists with PhDs, and even Nobel Prize winners. The phenomenon is especially prominent in STEM fields.

Two of the earliest systematic studies on Impostor Syndrome were Clance $\&$ Imes 1978 and Harvey 1981. The former conducted qualitative interviews with high-achieving individuals and their feelings of inadequacy, while the latter quantitatively measured Impostor Syndrome with Likert-scale statements. Both studies found that even talented and accomplished individuals can experience intense feelings of dis-belonging in high-achieving environments.

\section{Methods}

\subsection{Survey Structure and Motivation}

Surveys were given to undergraduate students at Pomona College taking Physics 71 (an introductory spring-semester physics course intended for physics majors) or Physics 42 (the introductory spring-semester physics course intended for non-majors). Two surveys were given to each class, one at the beginning of the semester and another several days after the students' received their grade for the first exam of the semester. Each survey contained Likert Sale statements, i.e. statements that students responded to on a scale from 1 to 5, with ``1" indicating strong disagreement and ``5" indicating strong agreement.

The first survey measured five dimensions:

\begin{enumerate}
    \item \textbf{Novice-like thinking}: eight questions from the Colorado Learning Attitudes about Science Survey (CLASS) were used in this survey. Student responses were compared to expert responses. Each completed survey was assigned a ``novice-like thinking score," defined as $N = \Sigma | a_n - e_n | $, where $N$ is the novice-like thinking score, $a_n$ is each of the student's answers, and $e_n$ is each of the experts' answers. A higher novice-like thinking score indicates a greater deviation from the way experts think, while a score of zero indicates complete alignment with the thinking style of experts.
    \item \textbf{Mindset}: four statements reflected a fixed mindset, while three statements reflected a growth mindset. Each completed survey was assigned a ``mindset score," defined as $M = \Sigma (5 - f_n) + \Sigma (g_n - 1)$, where $M$ is the mindset score, $f_n$ is each response to statements reflecting a fixed mindset, and $g_n$ is each response reflecting a growth mindset. A higher score indicates a more growth-oriented mindset, while lower scores indicate a fixed mindset.
    \item \textbf{Impostor Syndrome}: five statements reflected feelings of Impostor Syndrome, while three statements reflected belonging and confidence. Each survey response was assigned an ``Impostor Syndrome score, defined as $I = \Sigma (i_n-1) + \Sigma (5-b_n)$, where $I$ represents the Impostor Syndrome score, $i_n$ is each response to statements reflecting feelings of Impostor Syndrome, and $b_n$ is each response to statements reflecting feelings of belonging.
    \item \textbf{Workload Beliefs}: one statement reflected feelings of stress from the workload of the course, while three statements indicated that the course workload was manageable. Each completed survey was assigned a ``workload score," defined by $W = \Sigma(s_n-1)+\Sigma(5-m_n)$, where $W$ is the workload score, $s_n$ is the response to the stress-related statement, and $m_n$ is the response to the manageable-related statements. A larger Impostor Syndrome score indicates greater feelings of Impostor Syndrome. A larger workload score indicates greater feelings of anxiety or difficulty in managing the course workload.
    \item \textbf{Novice-like Studying}: nine statements reflected attitudes about effective ways of studying for an exam. (One statement directly referenced curriculum unique to the physics-major course, and thus it was removed from the survey for non-majors.) A survey containing these statements was provided to the professors of both courses. Each survey was assigned a ``novice-like studying score," defined as $S = \Sigma |p_n - s_n|$, where $S$ is the novice-like studying score, $p_n$ is each of the professor's responses, and $s_n$ is each of the students' responses. A lower novice-like studying score indicates that the student studied in a manner that aligns with the values of the professor.
\end{enumerate}

The second survey included the same questions regarding novice-like thinking, mindset, Impostor Syndrome, and workload beliefs. Respondents used pseudonyms for both surveys, preserving anonymity while allowing the surveyor to examine the changes in a students' responses. Each student was assigned a number indicating the change in novice-like thinking, change in mindset, and change in Impostor Syndrome, each equal to the score from the second survey minus the score from the first survey.

The novice-like studying questions were removed from the second survey. They were replaced with statements concerning the final dimension examined in this survey: \textbf{Academic Miscommunication (AM)}. Two statements indicated feelings of miscommunication between what the students had expected from the exam and the reality of the exam. Six statements indicated feelings of alignment between expectations and reality. Each student was assigned an ``Academic Miscommunication score," defined as $ A = \Sigma(m_n-1)+\Sigma(5-c_n)$, where $A$ is the Academic Miscommunication score, $m_n$ is each response to statements indicating miscommunication, and $c_n$ is each response to statements indicating accurate communication.

The survey also inquired about student demographics (such as race and gender), the students' academic background in physics, the students' motivation for enrolling in the course, and the students' attitudes about their relationship with their professor. The survey given to the professors also included questions about their academic values and attitudes toward their students.

\subsection{The Bonferroni Correction}

This study performed multiple hypothesis testing, exploring ten hypotheses across two classes, creating a total of twenty scatter plots. To avoid false positives, this study applies to Bonferroni correction, which requires that a p-value be equal to or lower than the standard p-value divided by the number of tests. With a standard p-value in psychology of less than or equal to 0.05, and with 20 tests, this study uses $p \leq 0.0025 $ as the threshold for statistical significance.

\subsection{Qualitative Interviews}

Six students across Physics 71 and 42 volunteered to participate in confidential, one-on-one interviews with the author. Students were not paid for their time, but they received chocolate afterwards as a gesture of gratitude. Interviews were conducted two weeks after students received their graded exams and took place in the Planetarium in the Estella physics building at Pomona College. Students signed a consent form prior to the interview. The audio of the interviews were recorded and transcribed using the software Temi; only the author had access to the audio and transcripts. The author obtained approval from the Pomona College Institutional Review Board before conducting the interviews.

The questions in the interviews probed five dimensions:

\begin{enumerate}
    \item \textbf{Academic Background}: students’ major, why they chose to enroll in physics, and whether they had taken a physics class before. 
    \item \textbf{Academic Miscommunication}: whether the students performed as well as they had hoped or expected, whether any questions surprised them, and whether the professor had provided them with an accurate impression of what the exam would consist of.
    \item \textbf{Mindset}: students were asked to provide an example of a time when they faced an especially challenging problem, a time when they felt accomplished, a time in which they received harsh and possibly unfair criticism. The students’ anecdotes can give insight into their mindset; for example, those with a growth mindset tend to feel accomplished when they overcome a difficult problem with great effort, while those with a fixed mindset feel accomplished when a problem is quick and effortless.
    \item \textbf{Impostor Syndrome}: their level of comfort in the physics department, such as raising their hand in class, studying in the physics lounge, attending office hours, etc. If time permitted, students were also asked whether they had ever felt misjudged by their peers.
    \item \textbf{Studying Methods}: students were asked to provide a detailed explanation of what methods they used to study for the exam, how early they studied for the exam, how many study sessions they engaged in, and how many hours they dedicated to studying.
\end{enumerate}

The qualitative interviews aimed to provide a deeper insight into students' attitudes. The students also provided possible explanations behind observed correlations in the data described in Section \ref{Results}.

\section{Survey Results}
\label{Results}

\subsection{Survey Results for Physics 42 (Non-Physics Majors) }

In Table \ref{Table2}, each row displays two variables being compared, the number of data points ($N$), the correlation coefficient $R^2$, the corresponding p-value, and whether we reject the null hypothesis.

\begin{table}[ht]
\begin{centering}
\begin{tabular}{|c|c|c|c|c|c|}
\hline
First Variable & Second Variable & N & R$^2$ & P-value & Reject Null? \\
\hline
Exam Scores & Novice-like Thinking & 16 & 0.285 & 0.0014 & Yes \\
Exam Scores & Change Novice-like Thinking & 28 & 0.014 & 0.54 & No \\
Exam Scores & Mindset & 33 & 0.006 & 0.67 & No \\
Exam Scores & Change in Mindset & 30 & 0.13 & 0.050 & No \\
Exam Scores & Impostor Syndrome & 35 & 0.258 & 0.0019 & Yes \\
Exam Scores & Change in Impostor Syndrome & 28 & $\approx$ 0.0 & $\approx$ 1 & No \\
Novice-Like Thinking  & Impostor Syndrome & 30 & 0.456 & 0.000042 & Yes \\
Exam Scores & Workload Beliefs & 34 & 0.14 & 0.029 & No \\
Exam Scores & Novice-Like Studying & 33 & 0.11 & 0.059 & No \\
Exam Scores & Academic Miscommunication & 30 & 0.457 & 0.000041 & Yes \\
\hline
\end{tabular}
\caption{For the physics course targeted to non-majors, high performance on the exam was correlated with expert-like thinking, low levels of Impostor Syndrome, and low levels of academic miscommunication. Higher levels of Impostor Syndrome also correlated with more novice-like thinking.}
\label{Table2}
\end{centering}
\end{table}

For each row in the table, the author produced a scatter plot comparing the two variables. Four of the ten scatter plots produced a statistically significant (p$\leq0.0025$) linear correlation:
\begin{enumerate}
    \item Exam scores inversely correlated with novice-like thinking. In other words, the students whose view on physics aligned with those of experts performed better on the exam than their peers.
    \item Exam scores inversely correlated with Impostor Syndrome. In other words, students with greater feelings of belonging performed better on the exam than their peers.
    \item Novice-like thinking directly correlated with feelings of Impostor Syndrome. In other words, students whose view on physics aligned with experts showed greater feelings of belonging than their peers.
    \item Exam scores inversely correlated with Academic Miscommunication. In other words, students whose expectations of the exam aligned with the reality of the exam performed better on the exam than their peers.
\end{enumerate}

We therefore reject the null hypothesis for these four pairs of variable. The scatter plots are shown in Figures \ref{graph1}, \ref{graph2}, \ref{graph3}, and \ref{graph4} below.

However, we fail to reject the null hypothesis for the other six pairs of variables ($p > 0.0025 $).

\begin{figure}[ht]
    \centering
    \begin{subfigure}[t]{0.49\textwidth}
        \centering
        \includegraphics[width=\linewidth]{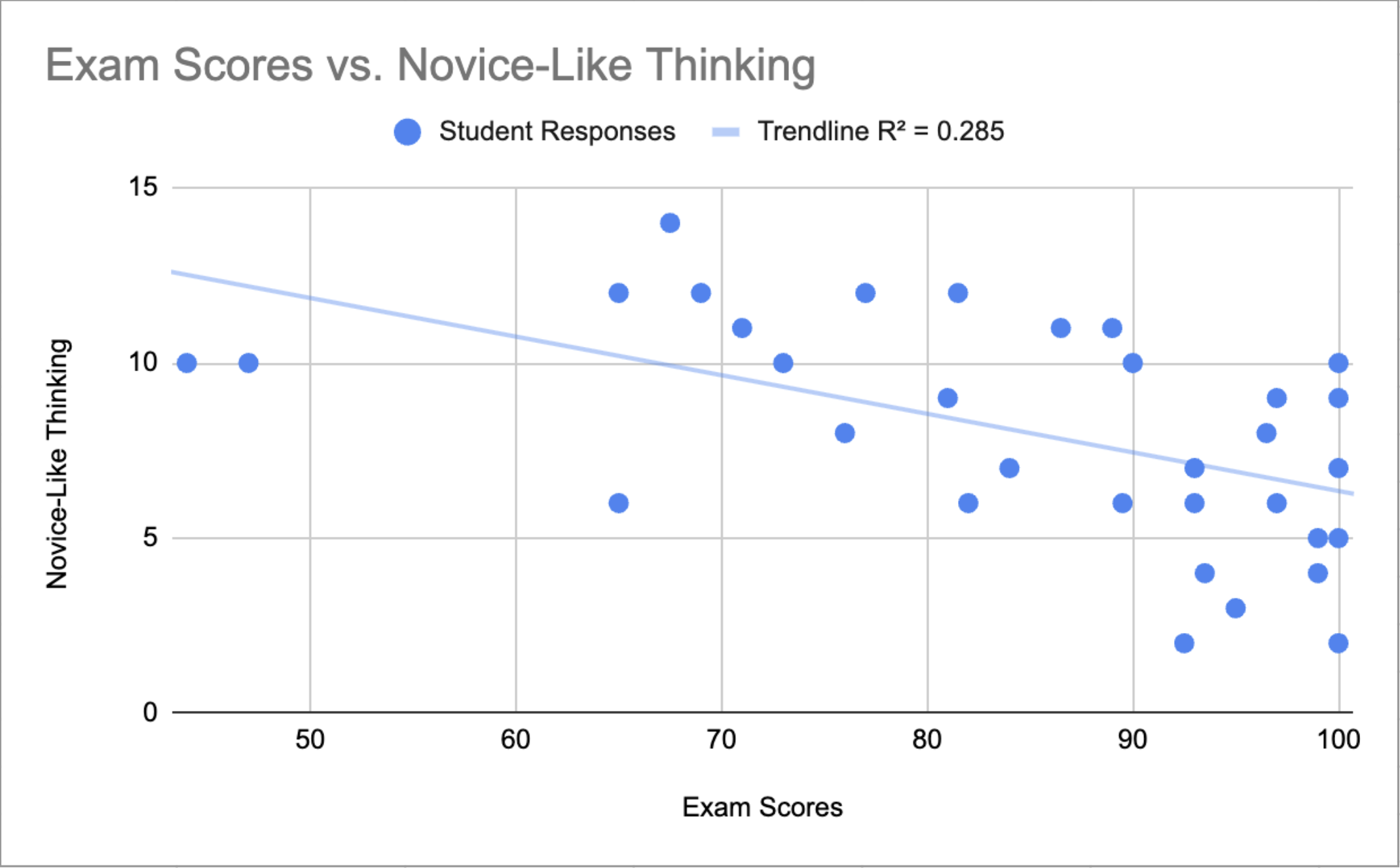}
        \caption{Students who performed well on the exam responded to the CLASS statements in a way that aligned with the responses of physics experts.}
        \label{graph1}
    \end{subfigure}
    \hfill
    \begin{subfigure}[t]{0.49\textwidth}
        \centering
        \includegraphics[width=\linewidth]{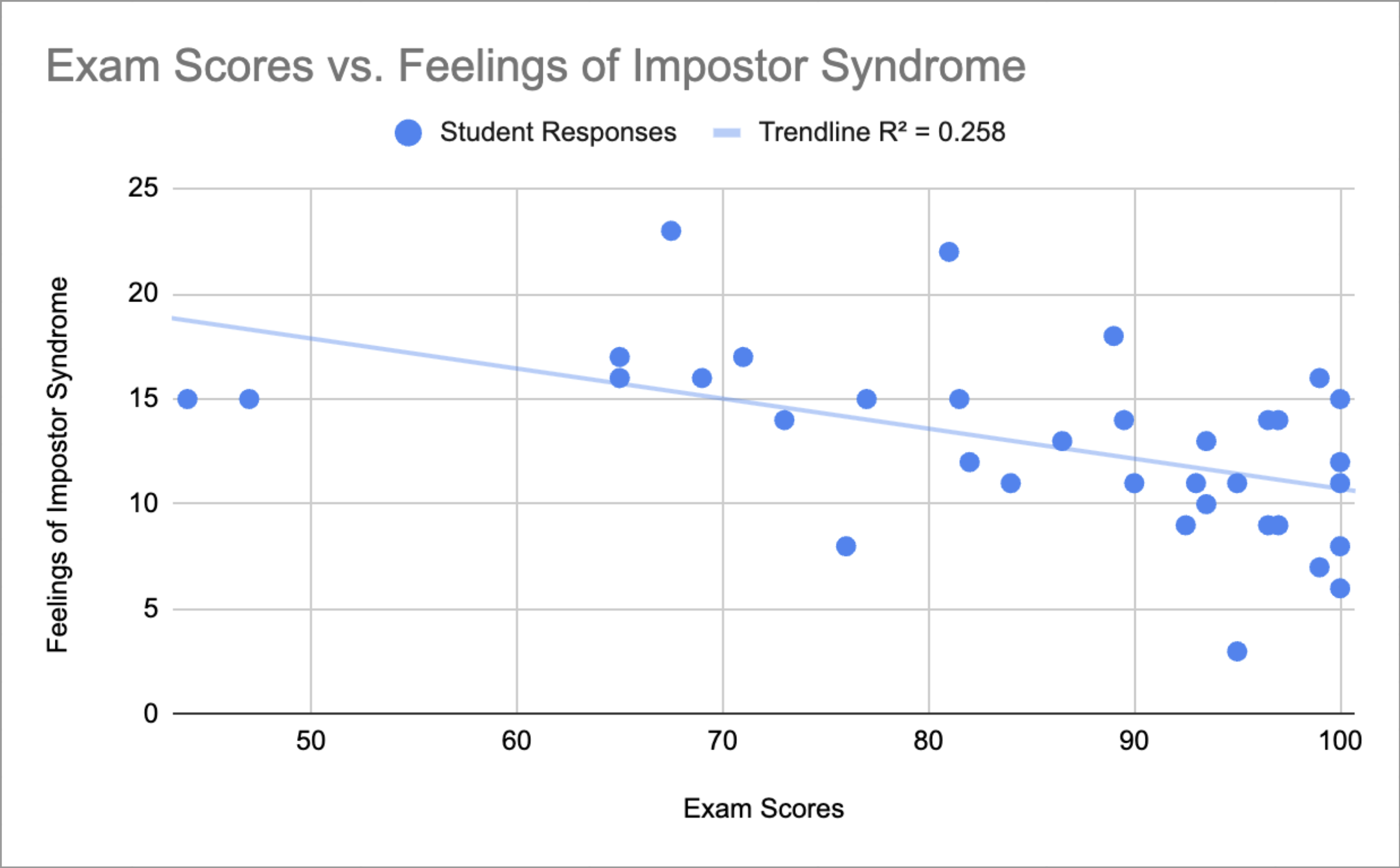}
        \caption{Students who performed poorly on the exam experienced greater levels of dis-belonging and inadequacy compared to their peers.}
        \label{graph2}
    \end{subfigure}
    \label{fig:two_graphs}
    \caption{1a (left) compares exam scores and novice-like thinking; 1b (right) compares exam scores and feelings of Impostor Syndrome.}
\end{figure}

\begin{figure}[ht]
    \centering
    \begin{subfigure}[t]{0.49\textwidth}
        \centering
        \includegraphics[width=\linewidth]{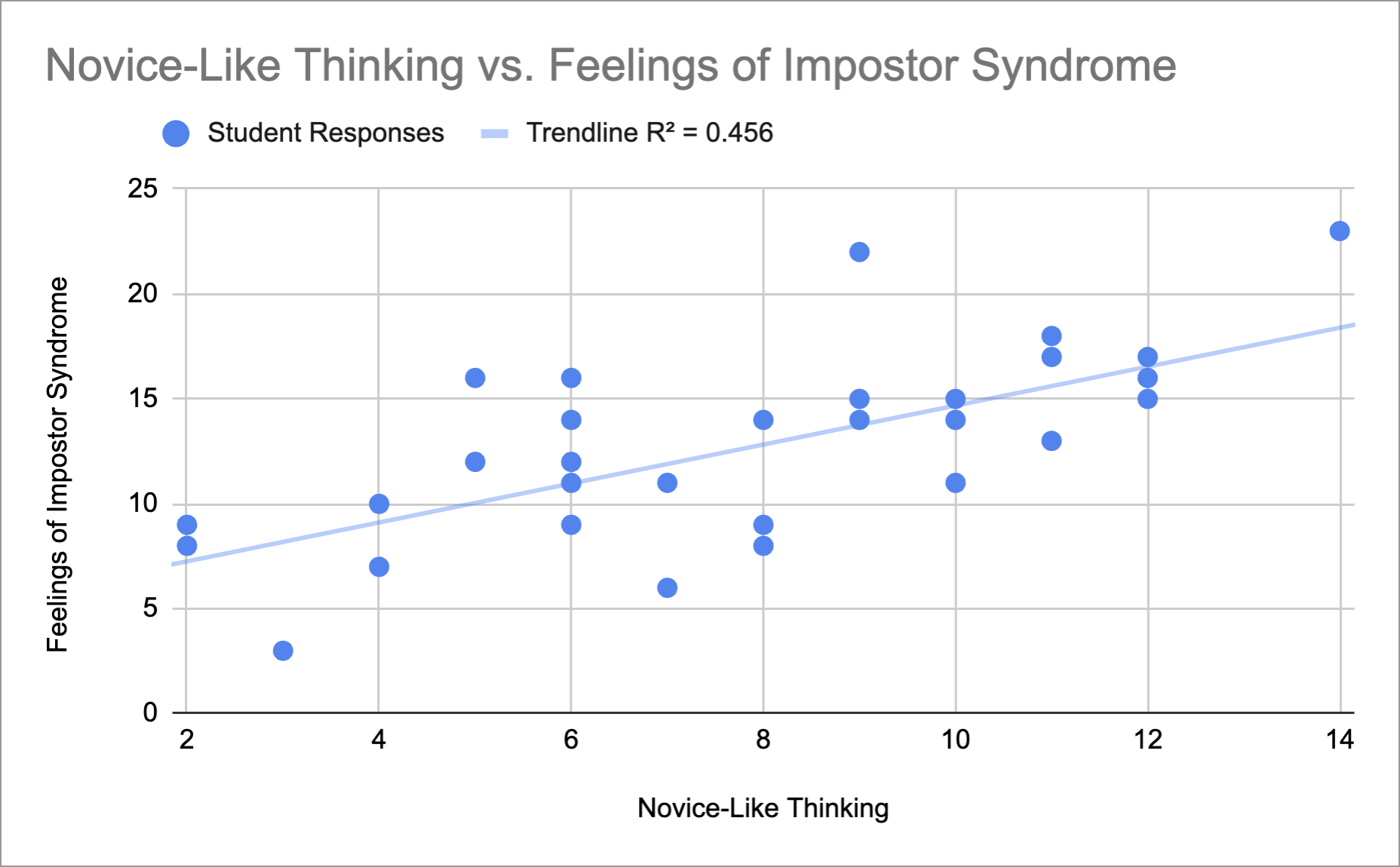}
        \caption{Students whose responses to CLASS aligned with experts' responses were more likely to feel that they belonged in the physics classroom.}
        \label{graph3}
    \end{subfigure}
    \hfill
    \begin{subfigure}[t]{0.49\textwidth}
        \centering
        \includegraphics[width=\linewidth]{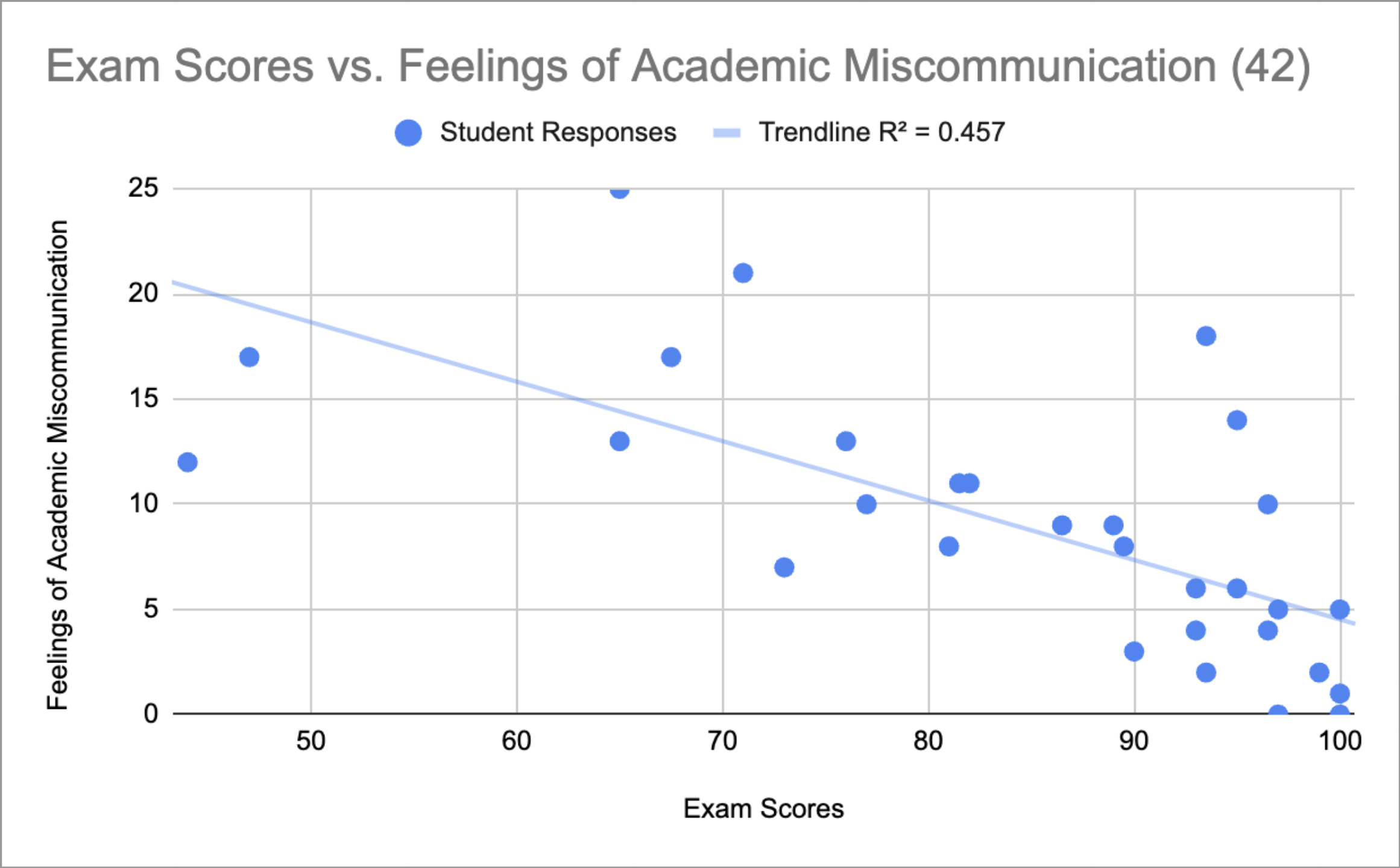}
        \caption{Students who performed poorly on the exam felt that they had miscommunicated with the professor about the form and content of the exam.}
        \label{graph4}
    \end{subfigure}
    \label{fig:two_graphs2}
    \caption{2a (left) compares novice-like thinking and feelings of Impostor Syndrome; 2b (right) compares exam scores and feelings of Academic Miscommunication.}
\end{figure}

\subsection{Survey Results for Physics 71 (Physics Majors)}

Table \ref{Table1} displayed the data pertaining to Physics 71. None of the ten scatter plots produced from the data showed a statistically significant linear correlation. Therefore, we fail to reject the null hypothesis for all ten pairs of variables.

\begin{table}[ht]
\begin{centering}
\begin{tabular}{|c|c|c|c|c|c|}
\hline
First Variable & Second Variable & N & R$^2$ & P-value & Reject Null? \\
\hline
Exam Scores & Novice-like Thinking & 16 & 0.24 & 0.054 & No \\
Exam Scores & Change Novice-like Thinking & 12 & 0.04 & 0.53 & No \\
Exam Scores & Mindset & 18 & 0.073 & 0.28 & No \\
Exam Scores & Change in Mindset & 13 & 0.40 & 0.020 & No \\
Exam Scores & Impostor Syndrome & 17 & 0.20 & 0.072 & No \\
Exam Scores & Change in Impostor Syndrome & 13 & 0.24 &  0.089 & No \\
Novice-Like Thinking  & Impostor Syndrome & 18 & 0.35 & 0.0097 & No \\
Exam Scores & Workload Beliefs & 18 & 0.03 & 0.49 & No \\
Exam Scores & Novice-Like Studying & 18 & 0.004 & 0.80 & No \\
Exam Scores & Academic Miscommunication & 14 & 0.29 & 0.047 & No \\
\hline
\end{tabular}
\caption{None of the pairs of variables displayed a correlation at or below the threshold for statistical significance.}
\label{Table1}
\end{centering}
\end{table}

\section{Interview Results}

This section explores the causes behind the four correlations presented in Section \ref{Results} by examining relevant excerpts from the qualitative interviews. The six students were given the following pseudonyms: Miranda, Ferdinand, Rosalind, Orlando, Viola, and Sebastian. Excerpts from the students that the author found to be particularly relevant are in bold. 

\subsection{Novice-like Thinking}

A recurring theme among the students was that rote memorization had proven to be an effective studying strategy in the past. However, the physics courses at Pomona College asked questions that tested their deep conceptual understanding, and rote memorization did not prepare them for these conceptual questions. In particular, students felt comfortable with each idea discussed in class individually, but they were unprepared to make nuanced connections between the ideas. As Viola explained:

\begin{quote}
“I always think the exams are very fair because they're always from information that I absolutely know will occur. [The professor] never says it explicitly, but there’s so many repetitions in our homework sets and what we do in class that I expected those questions.

I think the problem for me is that \textbf{every once in a while, the professor puts a little bit of a twist in the problems that requires you to connect conceptual things that she stated in class. For me, it’s a struggle to get those questions right. I always tend to get the questions that are very straightforward.}

We’ve already pretty much done an exact replica of it, either in the homework or the problems. But for the ones that were a little bit different, I’m not very good at them because I think I’m a little bit weaker, conceptually. Additionally, the professor always adds five questions in the beginning of the test that are supposed to be quite short, but they kind of flex your conceptual learning. And I always do poorly on those questions.”
\end{quote}

Several of the students echoed this sentiment — they felt comfortable with each idea on its own, but they struggled with questions that connected two or more familiar ideas in a novel manner.

Sebastian experienced a similar phenomenon, in which he understood each concept in the abstract, but he had difficulty with using the concept in a real-world situation, saying:

\begin{quote}
“I think it's awesome that physics—and my math classes as well—give a lot of partial credit. In my opinion, \textbf{I do know a lot of the concepts that are taught, but it’s difficult sometimes to apply them in practice.}”
\end{quote}

Ferdinand explained that physics required a different style of thinking compared to other college-level subjects, but that the overall enjoyed physics. He stated:

\begin{quote}
“I've been surprised at how much I liked it [physics]. \textbf{I think physics is a very different way of thinking than a lot of other subjects. For the first semester, it definitely took me a while to adjust to that way of thinking.} But I think now I very much honestly enjoy a lot of the topics that we learn about.

On the first exam [last semester], I got below a passing grade. It was atrocious, and I’ve never failed a college exam before. \textbf{I thought I understood the concepts, but then when it came to applying them, I struggled a little bit}, especially with time management because the exams can be really fast paced.”
\end{quote}

Generally speaking, students indicated that they entered the classroom with a “memorize and regurgitate” attitude, with the intention of remembering the meaning of vocabulary terms or equations and reciting the information on the exam. However, the homework and exams in the Pomona College physics department asked questions that required students to process the information and explore nuanced consequences that follow from the general principles learned in class. In other words, the classes require critical thinking and conceptual mastery. As a result, the students reported that they transitioned from a more novice-like memorize-and-regurgitate attitude to a more expert-like critical-thinking attitude over the course of the school year.

Viola shared an anecdote reflecting the transition from novice-thinking to expert-thinking:

\begin{quote}
“When you’re in cramming situation where you need to study for a test, suddenly you get into this mode of ‘Let me just memorize the equation and know how to apply it.’ But the thing is: sometimes you need to manipulate the situation. Sometimes the situation requires you to be flexible in thinking. \textbf{Having that rigid style of ‘I need to memorize the equation in order to answer the problem and plug in the numbers’ is not helpful in the long run.} And it took me a while to realize that in Physics 41 [the class prior to Physics 42].”
\end{quote}

The qualitative interviews suggest that novice-like thinking (e.g. memorizing information by rote) prevented students from performing well on the exam, while students who adopted expert-like thinking (e.g. deep conceptual understanding) were able to perform well on the exam, which is consistent with the linear correlation found in Section \ref{Results}.

Rosalind and Orlando both performed well on their exams. When asked about their studying methods, Rosalind stated: 

\begin{quote}
    “To study, I redid all the homework problems and all the activity sheet T-problems. \textbf{I did some T-problems that weren’t assigned}, but I didn’t do full-length problems [that weren’t assigned].”
\end{quote}

Orlando echoed this sentiment, saying: 

\begin{quote}
    “The morning of, I was sitting with two friends, and I was just going down [the list], doing like five T problems for every chapter—that kind of thing.”
\end{quote}

Generally speaking, the students who performed well on the Physics 71 exam also studied T-problems which were not explicitly assigned. This phenomenon reflects a general theme that emerged from the interviews: students who attempted to “learn the test” — that is, to learn information simply to restate it on the exam — performed poorly compared to students who aimed for mastery over the material itself.

\subsection{Impostor Syndrome}

The six students reported that they generally felt that they belonged in the classroom. They usually felt comfortable working with their classmates, attending office hours, or studying in the Estella physics building. The students exhibited warm feelings towards their peers, with some students stating that they had become close friends with their classmates. They also reported positive attitudes towards the physics professors, praising the professors’ approachability and availability. Overall, the levels of Impostor Syndrome were relatively low among the six interviewed students.

Despite students’ general comfort in the classroom, several students reported discomfort with raising their hand in class. Students with exam scores which were above the class average stated that they felt confident raising their hand in class. These students did not feel self-conscious or embarrassed asking questions, and they did not let their classmate’s perceptions of them hinder them from asking the questions which they were curious about. Ferdinand explained:

\begin{quote}
“I really like to ask questions in class. I think [the professor] does a really good job of making time for people to ask questions. \textbf{If there’s even one thing that I’m confused about, I'll ask a question, even if it’s not a fully formed question,} because what I found with physics is that things build on each other very quickly. \textbf{If you are confused about one topic, it is likely that you will be confused about it down the road when it comes up again.}

I think that, in the first semester, I was a little hesitant to ask questions sometimes. But this semester, I’m like, ‘I don't care what other people think of my question, because it’s a question that’s important to me.’ I also think that, if you have a question, it is likely that another person has the same question. I definitely feel very comfortable just as a whole in the classroom setting.”
\end{quote}

In contrast, students with exam scores which were below the class average expressed hesitation when asking questions in class. They were concerned that their classmates would already know the answer to their questions, and therefore they would consume valuable class time on a topic which was not beneficial to everyone in the classroom. Both students who expressed reservations with asking questions did not take physics in high school, and for that reason, they were concerned that their questions would be trivial or time-wasting for their peers who did take physics in high school. Sebastian explained:

\begin{quote}
    “It’s tough: sometimes I have questions for the professor that are specifically for me. And then that’s when I know I probably will not raise my hand during class and I will go to office hours. Those questions usually tend to be where I think it’s a basic concept that people may have prior knowledge coming into the class knowing. And it’s something that I don’t know. \textbf{It’s not that I’m afraid to ask—I just don't want to waste people's time.} So I’ll just go and ask him in office hours.”
\end{quote}

Viola expressed a similar view:

\begin{quote}
    “The professor always gives us opportunities to ask questions, but I'm always aware of the time limits. If I do ask a question, then that will take away from our lecture time—because it's so tightly scheduled. But she definitely does give us opportunities multiple times at office hours, of course. That's just something I'm always aware of. For me, I tend to shy away from asking questions in class.
    
    I don’t have a strong grasp on physics. \textbf{I don't know if my questions are basic questions where I’m just not aware of the concepts or if they are questions that are actually [helpful]. I don’t know. I don’t wanna spend other people’s time.} Time is so precious in physics. When we are in the lecture and I raise my hand, I feel like it has to be like a worthwhile thing to ask.”
\end{quote}

Self-consciousness was not the only phenomenon preventing students from raising their hand. Miranda explained that she takes time to form a question when the professor is lecturing, but that her classmates would form questions faster than she could. Thus, her classmates would raise their hands before she had fully formed her question, which interfered with her ability to ask questions of her own. Miranda elaborated:

\begin{quote}
    “Last semester, I was in Physics 70 [the class prior the Physics 71], and I think my section had around 30 people. Everyone was bursting with questions. So I was pretty intimidated. Generally, I don’t think fast enough. \textbf{It takes me like a long time to generate questions. But other people are always quicker than me. So at point I decided, ‘Okay, I could just listen to them and their questions to clarify whatever I was unsure about.’} I really learned the most in mentor sessions.

    But this semester I’ve been feeling okay. I feel like it’s because the material for 70 last semester was more novel to everyone. So naturally there were a lot of questions coming in. But this semester, it’s the basics of mechanics and Newton’s laws—not such exciting stuff. So it's generally the case that actually [the professor] asks questions and then we answer instead of us asking questions. I think it’s less frequent than last semester.”
\end{quote}

Miranda specified that, this semester, she feels much more comfortable raising her hand than in Physics 70. She said:

\begin{quote}
    “This semester, I’ve been more comfortable with raising my hand in the physics classroom. I think it’s because, firstly, I have a few friends in the class, so I'm more comfortable with the people generally. And secondly, it’s a smaller classroom.”
\end{quote}

Orlando also felt self-conscious at the beginning of the school year, but he became more comfortable as he developed friendships with his classmates:

\begin{quote}
    “At the beginning, a lot of the stuff that we were covering in class was going over my head. I have a pretty light physics background before this year. So I wasn’t super comfortable doing that yet. But the more time I spend in the department, the better I feel about it, I think, and right now, I feel very comfortable talking to people and being in mentor sessions and in class. \textbf{One of the reasons why I stuck with physics is because I made a lot of friends in my 70 class last semester. A lot of us are in 71 right now.}”
\end{quote}

Orlando’s experiences reflect a recurring theme from the interviews: students stay in physics because of the friends which they make during the first semester. It seems that the physics department thrives when it promotes a tight-knit and friendly community.

The link between comfort in raising one’s hand in the classroom and exam scores reflects the inverse correlation between exam scores and feelings of Impostor Syndrome displayed in Section \ref{Results}. This study is primarily correlational and leaves the causality undetermined. The author hopes that future work will investigate the following question: to what extent does a sense of belonging help the students perform well on exams, and to what extent does performing well on exams help the students gain a sense of belonging?

\subsection{Academic Miscommunication}

Two students, Sebastian and Miranda, reported feelings of miscommunicating with the professor about the content and style of the exam. Sebastian had assumed that the Physics 71 exam would consist of fewer, more long-form problems. However, the exam included a large number of shorter problem and one long-form problem. Sebastian explained:

\begin{quote}
    “When I take a test, I have no idea what the first one [in the semester] is going to look like. I chose to study a lot more of the word problems than the T-problems [two-minute problems]. \textbf{If I had known that the majority of the test was going be the T-problems, and that there would be one word problem, I would have chosen to spend my time studying more of the T-problems} than the Modeling [long-form] problems.”
\end{quote}

Sebastian was asked to elaborate on why he had believed that the test would be modeling-heavy and whether he felt that the professor had misguided him. Sebastian replied:

\begin{quote}
    “I just assumed that there would be more word problems. \textbf{It was not that [the professor] led me in any direction. It was more an assumption that I made based on what I know a test to be.}

    There was an activity sheet that the professor said would be really helpful for the test. I had assumed that the main thing on it was this word problem. And so I studied that and got that right on the test. But when I was given the test, I was like, ‘Oh, there are all these two minute problems that I could have spent the time to go over.’ They are super fast and fairly easy compared to a word problem, but there's more of them.”
\end{quote}

Sebastian correctly prepared for the topics that the exam covered, but he misunderstood the presentation style of the exam. He reported that he had studied thoroughly for the test, but he had studied the wrong style of question.

Importantly, the professor did not mislead Sebastian about what to expect on the exam. Instead, Sebastian relied on his previous assumptions about what a physics exam is supposed to be, but these assumptions led him astray.

Although Miranda performed well on her Physics 71 exam, she reported an experience of academic miscommunication on an exam in a different course in the Physics and Astronomy department at Pomona College, which this paper refers to as Course X to preserve its anonymity. She explained that she had implicitly assumed that the Course X exam would be mathematics-heavy, and therefore she spent much of her time creating a cheat-sheet filled with equations. However, the test was primarily conceptual. Miranda explained:

\begin{quote}
    “Generally, I find that doing a cheat sheet for all the exams is really helpful. […] \textbf{I spent probably three nights making a cheat sheet and it had a lot of diagrams and equations.} I feel more comfortable looking at mathematical equations and then trying to like understand things through equations. 

    \textbf{But then the [Course X] test was very conceptual—I don't think we had to use any equation at all.} […] So I think there’s a mismatch between what I expected to see on the test. There was more emphasis on the conceptual side of things and the very basic concepts. I think it requires you to get a very solid understanding of the basics.

But, I just didn't know how to prep for it. There was so much new information, and it seemed like everything was important. In the end, I think I was rather lucky to pass.”
\end{quote}

Miranda was asked if she felt that the professor and she had miscommunicated about the expectations for the exam. She replied:

\begin{quote}
    “Actually, \textbf{I don't know if it’s really miscommunication. I think I just didn't believe him when he said that!} I guess at one point he said that, ‘Okay, don’t litter your cheat sheet with equations,’ but I just didn't know like what else to do. I don't know what else to put on my cheat sheet.”
\end{quote}

Miranda misunderstood which style of thinking the exam questions would require. She studied for the test from a mathematical perspective, but the exam relied on deep conceptual understanding instead of mathematical prowess. As a result, the exam was unexpectedly difficult.

Sebastian and Miranda's anecdotes reflect a general theme that emerged in the interviews: students who performed poorly on the exam dedicated a similar amount of time and effort toward studying for the exam compared to their high-achieving peers. Those who struggled were no less hard-working than the other students; rather, they studied the wrong style of questions and adopted the wrong style of thinking in preparation for the exam. The professors did not mislead the students; rather, the students were misled by their prior expectations about science exams.

\section{Summary and Discussion}

This study explores an under-investigated problem in physics education: Academic Miscommunication (AM), the phenomenon in which a student develops beliefs about a class, its content, and its assignments that deviate from what the professor believed to have communicated. In Physics 42, an undergraduate physics course for non-majors at Pomona College, lower exam scores correlated with higher reported feelings of Academic Miscommunication (p=0.000041). In qualitative interviews, students explained that although they studied thoroughly for the exam, they misunderstood the layout of the exam or the style of problems. The professors did not give misleading instructions, but rather, the students had deep-seated assumptions about what a physics exam should consist of. These findings suggest that miscommunication of expectations may be a prevalent phenomenon in undergraduate physics courses, and that AM can prevent students from achieving their highest potential in class.

Higher exam scores in Physics 42 also correlated with attitudes about physics that aligned with those of physics experts (p=0.0014). The qualitative interviews suggest that physics novices adopt a ``memorize and regurgitate” attitude, learning each concept individually with the intent of repeating their knowledge on the exam. Physics experts, in contrast, are capable of finding subtle connections between ideas by engaging in critical thinking. High-performing students studied problems beyond what was assigned; they did not intend to “learn the test,” but rather, to learn the material.

Novice-like thinking correlated with higher feelings of Impostor Syndrome (p=0.000042). This study did not yield a definitive explanation as to the cause behind this correlation. However, many of the interviewed students shared that attending group study sessions and forging friendships with their classmates instilled them with a sense of belonging. The author hypothesizes that attending group study sessions caused students to learn better studying skills and adopt expert-like thinking while also lowering their feelings of Impostor Syndrome. The author encourages future researchers to explore this phenomenon.

Finally, higher exam scores correlated with lower feelings of Impostor Syndrome (p=0.0019). The author proposes two (compatible) hypotheses. Firstly, performing well in a course may cause a student to have a feeling of belonging in that field of study. Secondly, a student who feels that they belong in a field of study may be more likely to raise their hand in class, attend office hours, and attend group study sessions, causing a higher performance in the class. The author encourages future researchers to study this potential “chicken and egg” phenomenon.

\section{Acknowledgments}

Thank you to professors Natalie Mashian and Thomas Moore for their mentorship and feedback during the production of this project, and for allowing me to conduct research concerning them and their students.

Thank you to the Pomona College students who participated in the surveys and interviews.

Thank you to Niles Dorn and Christina Dong for their peer feedback in the early stages of this project.

\textit{Conflict of Interest Statement}: The author has no financial or non-financial conflicts of interest to disclose.

\textit{Generative AI Disclosure}: the Large-Language Model ChatGPT-5 was utilized in the process of re-formatting the bibliography to APA style; no AI models were used elsewhere in this paper.

\end{document}